\documentclass[conference,a4paper]{IEEEtran}
\usepackage{balance}
\usepackage{float}
\usepackage{standard}
\usepackage{etex}

\usepackage{rotate}
\usepackage{algorithmic}
\usepackage{algorithm}
\usepackage{float}
\usepackage{tikz}
\usepackage{refcount}

\usepackage{pgf,tikz}
\usepackage{pgfplots}
\usepackage{pgfplotstable}
\usepackage{bbm}
\usepackage{adjustbox}
\usetikzlibrary{calc}
\usepackage{relsize}
\usepackage{ifthen}

\usetikzlibrary{calc,fit,arrows,plotmarks,intersections,patterns,shapes,decorations,positioning, backgrounds, matrix}

\usepackage[numbers,sort&compress]{natbib}
\usepackage[ngerman, english]{babel}

\usetikzlibrary{arrows.meta}
\edef\wdArrowLength{2}
\tikzset{>={Latex[width=1.5mm,length=\wdArrowLength mm]}}
\usepackage{graphicx}

\usepackage{cancel}
\usepackage{mathtools}
\usepackage{shadethm}
\usepackage{empheq}
\usepackage{skull}
\usepgfplotslibrary{units}
\usepackage{calc}
\usepackage{nicefrac}
\usepackage{fancybox}
\usepackage{psfrag}
\usepackage{dsfont}

\title{Validating Properties of RIS Channel Models with Prototypical Measurements
\thanks{This work was supported in part by the German Federal Ministry of Education and Research (BMBF) in the course of the 6GEM Research Hub under grant 16KISK03. This project has received in part funding from the programme ”Netzwerke 2021”, an initiative of the Ministry of Culture and Science of the State of Northrhine Westphalia. The sole responsibility for the content of this publication lies with the authors. K. Weinberger, S. Tewes and A. Sezgin are with the Ruhr-Universit\"at Bochum, Germany Email:\{kevin.weinberger, simon.tewes, aydin.sezgin\}@rub.de.}} 
\author{Kevin Weinberger, Simon Tewes, Aydin Sezgin}
\date{\today}

\usepackage{graphicx}
\usepackage{placeins}

\usepackage{booktabs}
\usepackage{marvosym}
\usepackage{multirow}

\usepackage{latexsym, amsmath, amssymb, amsfonts, upgreek}

\usepackage[nolist]{acronym}

\usepackage{tikz}
\usetikzlibrary{calc,arrows,positioning,decorations,shadows,shapes,fadings,matrix}
\tikzset{>=latex'}
\tikzset{semithick}

\makeatletter
\providecommand{\IfElsePackageLoaded}[3]{\@ifpackageloaded{#1}{#2}{#3}}
\makeatother

\usepackage{subfigure}
\IfElsePackageLoaded{subfig}
	{\usepackage[subfigure]{tocloft}}{	
	\IfElsePackageLoaded{subfigure}
		{\usepackage[subfigure]{tocloft}}
		{\usepackage{tocloft}}
	}

\makeatletter

\def\tikz@delimiter#1#2#3#4#5#6#7#8{%
	\bgroup
		\pgfextra{\let\tikz@save@last@fig@name=\tikz@last@fig@name}%
		node[outer sep=0pt,inner sep=0pt,draw=none,fill=none,anchor=#1,at=(\tikz@last@fig@name.#2),#3]
		{%
			{\nullfont\pgf@process{\pgfpointdiff{\pgfpointanchor{\tikz@last@fig@name}{#4}}{\pgfpointanchor{\tikz@last@fig@name}{#5}}}}%
			\delimitershortfall\z@
			\resizebox*{!}{#8}{$\left#6\vcenter{\hrule height .5#8 depth .5#8 width0pt}\right#7$}%
		}
		\pgfextra{\global\let\tikz@last@fig@name=\tikz@save@last@fig@name}%
	\egroup%
}

\tikzset{hexagon/.code={
	\draw (0,2) -- (-4,0) -- (0,-2) -- (4,0) -- (0,2);
}}

\tikzset{phone/.code={
   \node [rectangle,rounded corners=1.5pt,draw,minimum height=0.6cm, minimum width=0.35cm] at (0,0){};
   \node [rectangle,rounded corners=1.5pt,draw,minimum height=0.5cm, minimum width=0.3cm] at (0,0){};
}}

\makeatletter
\def\cantox@vector#1#2#3#4#5#6#7#8{%
  \dimen@.5\p@
  \setbox\z@\vbox{\boxmaxdepth.5\p@
   \hbox{\kern-1.2\p@\kern#1\dimen@$#7{#8}\m@th$}}%
  \ifx\canto@fil\hidewidth  \wd\z@\z@ \else \kern-#6\unitlength \fi
  \ooalign{%
    \canto@fil$\m@th \CancelColor
    \vcenter{\hbox{\dimen@#6\unitlength \kern\dimen@
      \multiply\dimen@#4\divide\dimen@#3 \vrule\@depth\dimen@\@width\z@
      \vector(#3,-#4){#5}%
    }}_{\raise-#2\dimen@\copy\z@\kern-\scriptspace}$%
    \canto@fil \cr
    \hfil \box\@tempboxa \kern\wd\z@ \hfil \cr}}
\def\bcancelto#1#2{\let\canto@vector\cantox@vector\cancelto{#1}{#2}}
\makeatother

\newcommand{\ifthen}[2]{\ifthenelse{#1}{#2}{}}

\newcommand{\listequationsname}{List of Formulas}
\newlistof{myequations}{equ}{\listequationsname}

\definecolor{myblue1}{rgb}{0,0,255}
\definecolor{myblue2}{rgb}{65,105,225}
\definecolor{myblue3}{rgb}{70,130,180}
\definecolor{myblue4}{rgb}{176,196,222}

\newcommand{\mytilde}{{\raise.17ex\hbox{$\scriptstyle\mathtt{\sim}$}}}
\newcommand{\naive}{}
\def\naive/{na\"{\i}ve}

\newcommand{\executeiffilenewer}[3]{%
\ifnum\pdfstrcmp{\pdffilemoddate{#1}}%
{\pdffilemoddate{#2}}>0%
{\immediate\write18{#3}}\fi%
}

\pgfkeys{/pgf/fpu}
\pgfmathparse{16383+1}

\pgfkeys{/pgf/fpu=false}

\IEEEoverridecommandlockouts

\begin{document}
\bstctlcite{IEEEexample:BSTcontrol}

\maketitle

\begin{abstract}
  The integration of Reconfigurable Intelligent Surfaces (RIS) holds substantial promise for revolutionizing 6G wireless networks, offering unprecedented capabilities for real-time control over communication environments. However, determining optimal RIS configurations remains a pivotal challenge, necessitating the development of accurate analytical models. While theoretically derived models provide valuable insights, their potentially idealistic assumptions do not always translate well to practical measurements. This becomes especially problematic in mobile environments, where signals arrive from various directions. This study deploys an RIS prototype on a turntable, capturing the RIS channels’ dependency on the angle of incoming signals. The difference between theory and practice is bridged by refining a model with angle-dependent reflection coefficients. The improved model exhibits a significantly closer alignment with real-world measurements. Analysis of the reflect coefficients reveals that non-perpendicular receiver angles can induce an additional attenuation of up to $-14.5$dB. Additionally, we note significant phase shift deviations, varying for each reflect element.
\end{abstract}

	\vspace{-0.1cm}
\section{Introduction}
Reconfigurable Intelligent Surfaces (RISs) are envisioned as a pivotal technology for realizing the ambitious goals of future sixth-generation (6G) wireless communication networks \cite{RISBasar1}.  An RIS comprises an extensive array of passive elements, strategically configured to reflect electromagnetic signals, thereby reshaping the propagation characteristics of the wireless environment \cite{renzoAlessio}. This unique feature not only turns the RIS-assisted network into a facilitator for a diverse range of innovative applications at the physical layer \cite{outageResilience,UAV_Kev,SynBenConf}, but also enhances various communication goals, including capacity, resilience, and energy efficiency \cite{RISBasar2,EEJourn,SynBenConf}. In order to achieve these goals, the appropriate RIS configurations need to be determined. Due to the complexity of these problems, this is  usually done by solving specific optimization problems. To this end the precision of these problems is highly dependent on the utilized channel model \cite{EEJourn}. Additionally, effects that are present in the real-world, are often not accounted for in idealistic channel models. This is especially pronounced in mobile wireless environments, where the reflected waves of the RIS are often utilized to bypass blockages \cite{outageResilience,UAV_Kev,RISBasar2}. This means that the RIS is usually deployed to reflect incoming signals in a non-perpendicular angle, which inevitably leads to additional attenuation and imperfect phase shifts \cite{RIS_phaseDep,expValid1}.

The intention of this work is to experimentally refine the precision of a common analytical channel model often used in the state of the art literature \cite{RIS_phaseDep,KevChanMod,UAV_Kev}. More precisely, we aim to refine the analytical channel model by calculating and analyzing the angle-dependent reflection coefficients that are observable in real-world measurements. As a result, we are able to make the channel model significantly more precise with regards to the experimental measurements and bridge the gap between theory and practice. Further, valuable insights are drawn from the determined coefficients and highlight the significant impact that the reflection angle has on the RIS's practical efficacy.
\section{System and Channel Model}\label{sec:sysmod}
The system model outlined in this paper assumes a transmitter (Tx) with a single antenna transmitting towards the reconfigurable intelligent surface (RIS), comprising $M=256$ reflective elements in accordance with the prototype employed in the experimental configuration. Given the RIS's ability to apply configurable phase shifts to reflections, the effective Tx-RIS-Rx channel can be formulated as follows:
 \vspace{-0.1cm}
 \begin{align}\label{eq:heff}
    h^\mathsf{eff} =  \sum_{m=1}^M {h}_m \theta_m {g}_m, \\[-18pt] \nonumber
 \end{align}
  where ${h}_m$ (${g}_m$) denotes the channel coefficient between the transmitter (receiver) and the $m$-th RIS element. The phase shift, that is induced on the reflection by the $m$-th reflecting element, is defined as $\theta_m  =  A_m(\varphi_m)e^{j\varphi_m}$,
 where $\varphi_m$ is the adjustable phase and $A_m(\varphi_m) \in [0,1]$ the phase-dependent reflect amplitude.
 With the above considerations, the cascaded channel between Tx-RIS-Rx over element $m$ is written as  \vspace{-0.1cm}
 \begin{align}
     {h}^\mathsf{casc}_m = h_m g_m,
 \end{align}
In order to determine the  channels for every reflect element $m$ analytically, we intend to utilize the geometry of the setup. To this end, we denote $d^h_m$ ($d^g_m$) as the distance between the transmitter (receiver) and the $m$-th RIS element. According to \cite{goldsmith2005wireless}, the cascaded channel between Tx-RIS-Rx over the $m$-th
 RIS element is then given by
 \begin{align}\label{eq:chanModel}
     {h}^\mathsf{casc}_m = h_m g_m = \Big[ \frac{c}{4\pi f d^h_m} e^{j\frac{2\pi}{\lambda}d^h_m} \Big]  \Big[ \frac{c}{4\pi f d^g_m}  e^{j\frac{2\pi}{\lambda}d^g_m} \Big] ,
 \end{align}\\
 where $c$ is the speed of light, $f$ is the frequency and $\lambda$ is the wavelength of the signal. To verify this channel's precision, a real-world scenario employing an RIS prototype is constructed in the next section.
\section{Experimental Validation}\label{sec:exp}
\subsection{Experimental Setup}\label{subsec:expSetup}
\begin{figure}[ht]
	\centering
	\includegraphics[width=.75\linewidth]{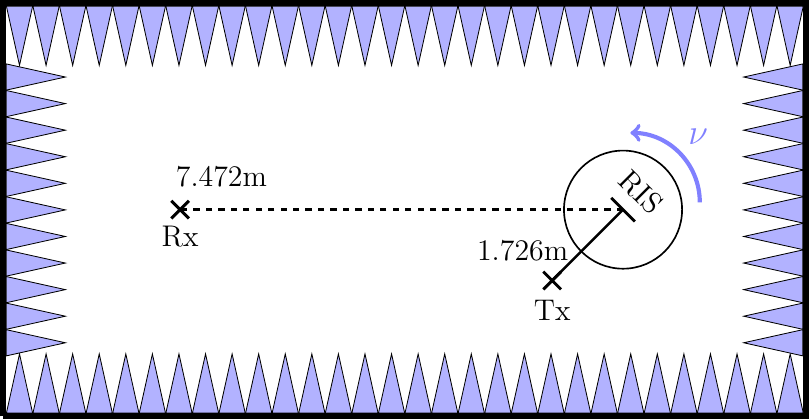}
	\caption{Top view of the anechoic chamber, with the antenna positions at the respective distances and angles marked with crosses. The setting shown is for an RIS angle $\nu=45^\circ$.}
	\label{fig:Skizze}
\end{figure}

\begin{figure}[ht]
	\centering
	\includegraphics[width=0.95\linewidth]{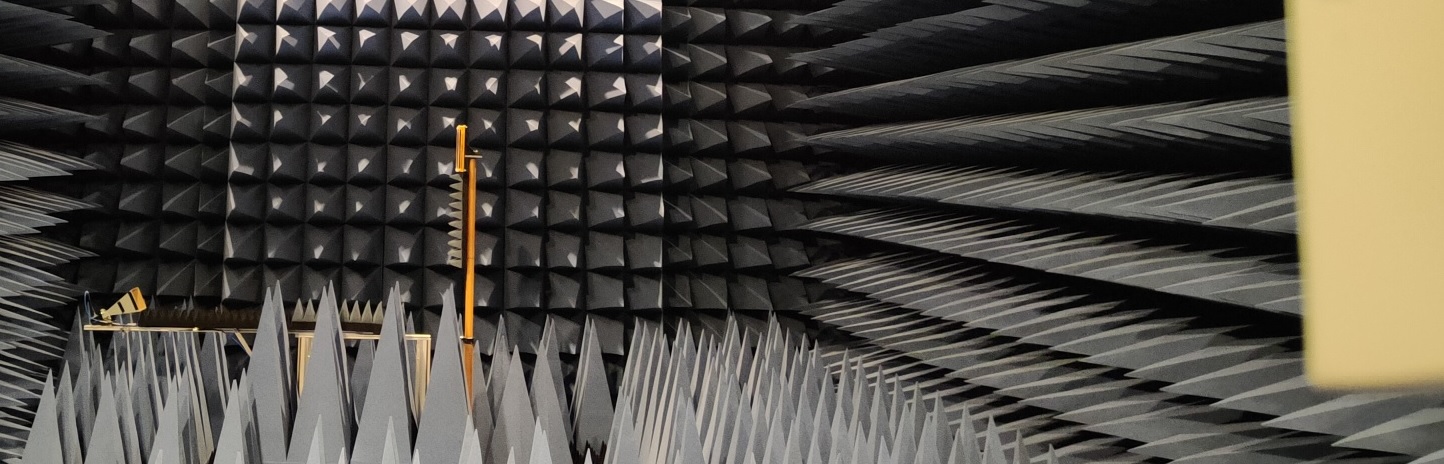}
	\caption{Photo of the RIS and antennas in the anechoic chamber.}
	\label{fig:Setup}
\end{figure}

To validate the precision of the channel model described in section~\ref{sec:sysmod} we set up a scenario utilizing an RIS prototype in the 5GHz band. The RIS prototype used has an overall size of 360 by 247 mm and is described in greater detail in \cite{ISAPpaperMH}.
A total of 256 elements layed out in a 16 by 16 grid are available, where each element features an RF switch to achieve the phase shifting. The surface is designed as a binary-switching surface with a $180^\circ$ phase shift at the designed carrier frequency of $f^\mathsf{RIS} = 5.53$GHz, i.e., a wavelength of $\lambda^\mathsf{RIS}=5.42$cm. To estimate the path loss across the effective RIS channel $h^\mathsf{eff}$ we use a vector network analyzer of type PicoVNA 106 from manufacturer PicoTech. The S21 parameter is evaluated over a span of 200 MHz, from 5.4 GHz to 5.6 GHz with an resolution bandwidth of 10 kHz. Each of the two ports of the VNA is connected to a directional horn antenna of type LB-187-15-C-SF from manufacturer A-Info. The gain of the antennas within the chosen span is at least 16.89 dBi. The transmit power is set to 0~dBm. To utilize the dynamic range of the VNA optimally an LNA is used at the receive antenna to counteract the path loss of the channel. Here, the LNA of type Mini-Circuits ZX60-153LN-S+ has a gain of {16.92}{dBm} at {5}{GHz}. To reduce the effect of multipath all measurements were conducted within an anechoic chamber. The chamber used is a fully-anechoic chamber of dimension {7}{m} x {13.5}{m} x {6}{m} from the manufacturer Frankonia.

The top view of the geometry of the setup is depicted in Fig. \ref{fig:Skizze}. Moreover, the RIS was placed on a tripod at a height of 2.26m centrally on the turntable embedded in the floor. The receiving antenna was placed at the same height at a distance of 7.472m from the RIS, so that for a turntable angle $\nu$ of $0^\circ$, the receiving antenna is aligned perpendicular to the RIS. The angle of the receiving antenna to the RIS thus always corresponds to the turntable angle. The transmitting antenna was placed on an extended antenna bracket on the turntable below the RIS. The transmitting antenna was located at a height of 1.36m, resulting in a total distance of 1.96m from the center of the RIS. The vertical angle of incidence to the RIS is therefore $30.6^\circ$. Due to the placement of the transmitting antenna on the turntable with the RIS, it is static to the RIS when it is rotated and illumination of the RIS always takes place at a horizontal angle of $0^\circ$. A picture of the setup is shown in Fig.~\ref{fig:Setup}, in which the Rx antenna is seen on the right, while the Tx antenna is seen on the left, slightly tilted upwards. Due to this arrangement and tilting of the transmitting antenna, the specular reflection case does not occur.

Note: In the perpendicular case at an rotating stage angle of $0^\circ$, the transmitting and receiving antennas are in line, so that the side lobes of the transmitting antenna would radiate back towards the receiving antenna. To prevent this, the transmitting antenna was additionally enclosed with absorber material to attenuate the side lobes.

\subsection{Model-based Optimization}
To refine the channel model specified in section \ref{sec:sysmod} with the prototype, we determine an optimized RIS configuration using the baseline channel model, which is based on the established geometry. By applying these configurations in our measurements, we are able to quantitatively compare the channel values of the experimental-based RIS setup with the ones determined by the suggested channel model. We introduce angle-dependent reflect coefficients, intended to model the difference between theory and practice, and are able to determine the properties that are missing in our baseline channel model.

In more details, we compute the optimal RIS configuration analytically, expressed as $\varphi_m^*(C_t) = C_t - \varphi'_m$, where $C_t$ denotes an arbitrary value for the desired phase at the receiver, taking into account the absence of a direct Tx-Rx link, and $\varphi'_m = \frac{2\pi}{\lambda^*}(d_m^h + d_m^g)$. Given the flexibility in selecting $C_t$, we iterate through $T = 360$ evenly-distributed phase values $C_t$, rounding the continuous phases to the nearest binary switching state of the RIS prototype. Subsequently, we are able to assess and compare the simulated performances of the resulting RIS-facilitated links. This process can be represented mathematically as \vspace{-0.4cm}

\begin{align}
h_t^\mathsf{eff} = &\hspace{-0.1cm} \sum_{m=1}^M {h}^\mathsf{casc}_m A_m(\text{rd}(\varphi_m^*(C_t))) \e^{j \text{rd}(\varphi_m^*(C_t))},\,  \forall C_t \in \mathsf{\Theta},\label{optEq}\\
 \text{with}&  \quad A_m(\tau) = \begin{cases} $0.5012 (-3\text{dB})$ , \quad\text{if } \tau = \pi \\ 1 , \quad\quad\quad\quad\quad\,\, \,\,\,\,\:\text{otherwise}\end{cases}\ \,\,\,\, ,\\
 &\quad\,\,\,\, \text{rd}(\tau) = \begin{cases} \pi , \quad\text{if } \frac{\pi}{2} \leq\tau < \frac{3\pi}{2} \\ 0, \quad\text{otherwise}\end{cases}\quad\quad\quad\,\,\,\,\,\,  , \label{rdFun}\end{align}\begin{align}	
 &\quad\quad\quad\mathsf{\Theta} = \left\{0, \frac{\pi}{180}, \frac{2\pi}{180}, \frac{3\pi}{180}, \ldots, 2\pi - \frac{\pi}{180} \right\} ,\label{Theta} \\[-13pt] \nonumber
\end{align}	
where we assume an additional attenuation of 3dB to the reflected path of reflecting element $m$ if it is turned on, capturing the hardware limitations of the utilized RIS prototype \cite{ISAPpaperMH}. Note that in addition to the set of optimal phase values for each $C_t$ from (\ref{optEq}), the respective analytical channel model is also determined. Moreover, this procedure can be used for optimization towards any desired reflection angle because only $h_m^\mathsf{casc}$ is dependent on the geometry of the system.

\subsection{Refining the Channel Model}

In order to capture the effects of the angle-dependency on the RIS reflection, we refine our model by introducing an angle-dependent value $\alpha_m(\nu)$ into our channel model as
\begin{align}\label{eq:alpha}
\tilde{\theta}_m  = \alpha_m(\nu)  \underbrace{A_m(\varphi_m) e^{j\varphi_m}}_{\theta_m},
\end{align}
where $\nu$ is the relative angle of the Rx as visualized in Fig. \ref{fig:Skizze} and $\alpha_m(\nu)$ is the angle-dependent complex reflect coefficients of RIS element $m$ with a magnitude less than or equal to one, i.e., $\{ \alpha_m(\nu) \in \mathbb{C} : |\alpha_m(\nu)| \leq 1 \}$.

With the use of the experimentally recorded measurements, belonging to the respective analytically determined configuration, we can formulate a system of linear equations $\mat{H}\vect{\alpha} = \vect{y}$ as
\begin{align}\label{eq:linEq}
\begin{bmatrix}
   h_{1}^\mathsf{casc}{\theta} _{1,1} & \ldots & h_{M}^\mathsf{casc}{\theta}_{M,1} \\
   \vdots & \ddots &  \vdots &\\
   h^\mathsf{casc}_{1}{\theta}_{1,T} & \ldots &  h^\mathsf{casc}_M \theta_{M,T}
\end{bmatrix}
\begin{bmatrix}
   \alpha_1(\nu) \\
   \vdots \\
   \alpha_M(\nu)
\end{bmatrix}=
\begin{bmatrix}
   y_1\\
   \vdots\\
   y_T
\end{bmatrix},
\end{align}
where $y_t$ is the measured channel value using the configuration corresponding to the phase offset $C_t$. Due to the fact that we utilize $T=360$ phase offsets, the matrix $\mat{H}$ should be overdetermined and the problem solvable by using the Least Squares Method, i.e., $\vect{\alpha} = (\mat{H}^T \mat{H})^{-1} \mat{H}^T \vect{y}$. However, due to the quantization of the RIS phase shifts in (\ref{rdFun}), it becomes possible for configurations of neighbouring phase offsets $C_t$ to share the same RIS configuration. Hence, we also include a single element sweep into our measurement campaign, guaranteeing 256 distinct RIS configurations. To further improve the quality of the calculated reflect coefficients, we repeat the same procedure using a tiling approach, where 4 neighbouring reflect elements are grouped and treated as a single element, adding another 64 measurements. Overall, we determined the respective 360 configurations optimizing towards $\nu^\mathsf{opt} = \{0^\circ,0.1^\circ,1^\circ,2^\circ,10^\circ,30^\circ,45^\circ,60^\circ\}$ as well as their tiled versions. This was achieved by changing the geometry of the system, i.e., $h_m^\mathsf{casc}$ during simulations in (\ref{optEq}). By denoting $\tilde{\mat{H}_\mu} = [\mat{H}_\mu^T , \mat{H}_{\mu,\mathsf{tiled}}^T]^T$ and $\tilde{\vect{y}_\mu} = [\vect{y}_\mu^T , \vect{y}_{\mu,\mathsf{tiled}}^T]^T$, where $\mu$ is the angle that the RIS is optimized towards for the $T=360$ configurations as well as their tiled versions, we can express the overall system of linear equations as
\begin{align}
 \hat{\mat{H}} \vect{\alpha} = \hat{\vect{y}},
\end{align}
where $\hat{\mat{H}} = [\mat{H}_\mathsf{single}^T ,\mat{H}_\mathsf{single,tiled}^T, \tilde{\mat{H}}_{0^\circ}^T ,\tilde{\mat{H}}_{0.1^\circ}^T , \dots, \tilde{\mat{H}}_{60^\circ}^T]^T$ is a stacked matrix with the analytical values and $\hat{\vect{b}} = [\vect{y}_\mathsf{single}^T ,\vect{y}_\mathsf{single,tiled}^T, \tilde{\vect{y}}_{0^\circ}^T ,\tilde{\vect{y}}_{0.1^\circ}^T , \dots, \tilde{\vect{y}}_{60^\circ}^T]^T$ is a stacked vector with the experimental measurements depending on $\nu^\mathsf{opt}$.

\section{Experimental Results and Parameter Fitting}
\subsection{Analysis of Beampattern and Refined Channel Precision}
\begin{figure}
\centering
\includegraphics[width=.95\linewidth]{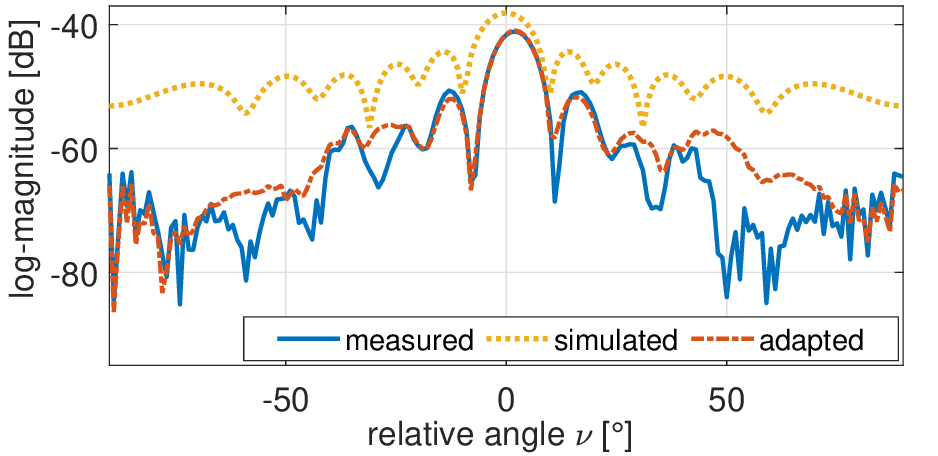}
\caption{The logarithmic magnitude of the channel values $h^\mathsf{eff}$ measured, simulated (using $\theta_m$) and adapted (using $\tilde{\theta}_m$) for a determined RIS configuration reflecting towards $0^\circ$.}
\label{fig:0_deg_BF}
\end{figure}
\begin{figure}
\centering
\includegraphics[width=.95\linewidth]{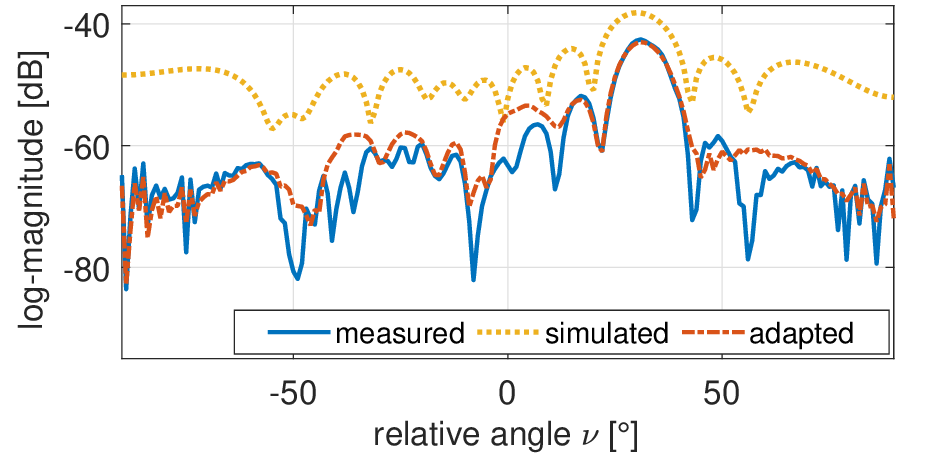}
\caption{The logarithmic magnitude of the channel values $h^\mathsf{eff}$ measured, simulated (using $\theta_m$) and adapted (using $\tilde{\theta}_m$) for a determined RIS configuration reflecting towards $30^\circ$.}
\label{fig:30_deg_BF}
\end{figure}
\begin{figure}
\centering
\includegraphics[width=.95\linewidth]{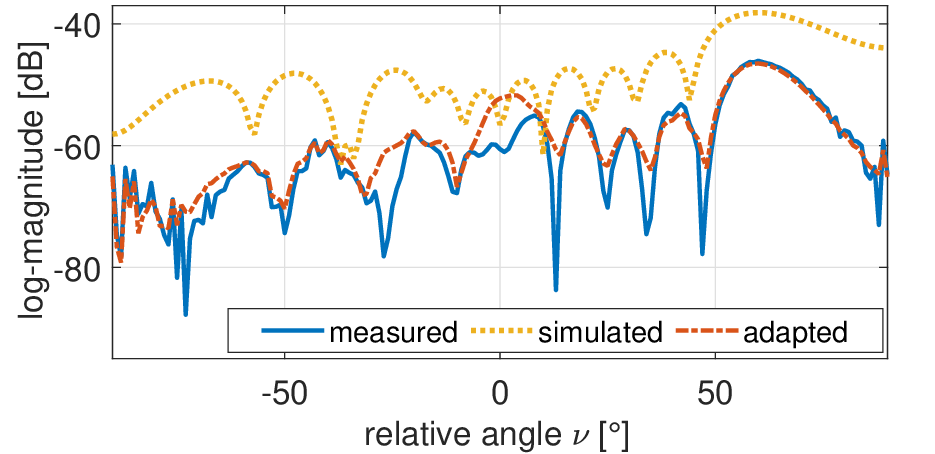}
\caption{The logarithmic magnitude of the channel values $h^\mathsf{eff}$ measured, simulated (using $\theta_m$) and adapted (using $\tilde{\theta}_m$) for a determined RIS configuration reflecting towards $60^\circ$.}
\label{fig:60_deg_BF}
\end{figure}
In Fig. \ref{fig:0_deg_BF}, we visualize the beampattern for the surface, configured to reflect towards $0^\circ$, for the measured case, simulated case without using the determined reflect coefficients $\alpha_m(\nu)$, i.e., $\theta_m$, and adapted case, where $ \alpha_m(\nu)$ is utilized, i.e., $\tilde{\theta}_m$. More precisely, Fig. \ref{fig:0_deg_BF} shows that the measured as well as the simulated beampatterns have a pronounced peak at $0^\circ$. The figure also shows that the shape of the sidelobes that are visible in the simulations can also be seen in the measurements. However, the overall measurements seem to be stronger attenuated, which gets more pronounced the higher the relative angle towards the receiver is set. It can be seen that this effect, which is not accounted for in the simulations, is considered when re-simulating the scenario utilizing the calculated $\alpha_m(\nu)$. The beampattern of the adapted channel values therefore shares a high resemblance with the measured one, with the highest precision at the main beam and its first order side lobes. A similar relation between the measured and simulated results can be observed in Fig. \ref{fig:30_deg_BF} for a RIS configuration steering the beam towards $30^\circ$. Interestingly, in this figure the adapted beamplot is deviating from the measured beamplot at $\nu = 0^\circ$ and $\nu=60^\circ$. Therefore, the utilized set of RIS configurations determined with (\ref{optEq}) seems to introduce a bias into the determined $\alpha_m$ at $\nu=0$ and $\nu = 60^\circ$, which may be introduced due to using a set of 360 optimal configurations towards this direction. The same effect can also be observed at $\nu=0$ in Fig. \ref{fig:60_deg_BF}, where the surface is configured to reflect towards $60^\circ$.
\subsection{Analysis of the Angle-dependent Reflect Coefficients}
\begin{figure}
\centering
\includegraphics[width=.975\linewidth]{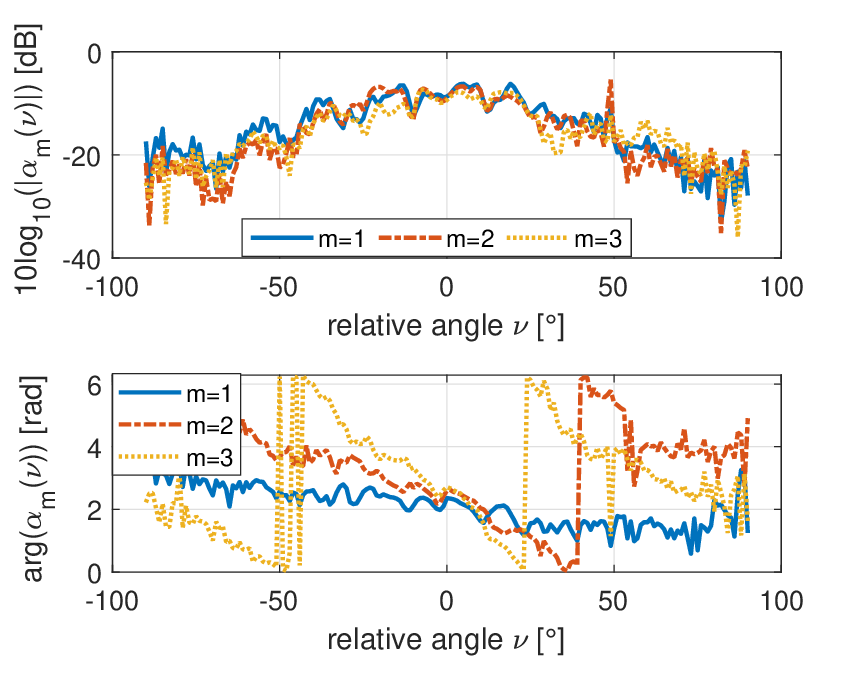}
\caption{Logarithmic magnitude (top) and phase (bottom) of the angle-dependent reflect coefficient $\alpha_m(\nu), \forall  m\in\{1,2,3\}$}
\label{fig:123}\vspace{-0.5cm}
\end{figure}

\begin{figure}
\centering
\includegraphics[width=.975\linewidth]{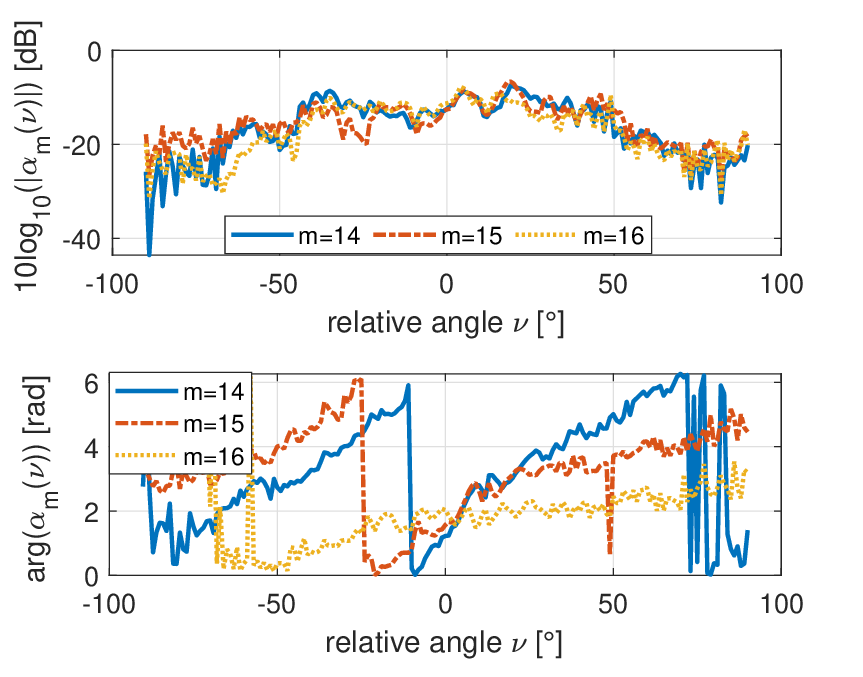}
\caption{Logarithmic magnitude (top) and phase (bottom) of the angle-dependent reflect coefficient $\alpha_m(\nu), \forall  m\in\{14,15,16\}$}
\label{fig:141516}\vspace{-0.6cm}
\end{figure}

In order to get a better understanding of the angle-dependent reflect coefficients $\alpha_m(\nu)$, this section aims to analyze their impact by evaluating the induced magnitude and phase differences. To this end, we calculate $\alpha_m(\nu)$ as defined in (\ref{eq:linEq}), but only use the single-element RIS configurations, including the tiled versions, in order to remove potential biases, which were observed when also using the optimized configurations. We define a row-wise indexing of the surface, starting from the top left corner and ending in the top right corner before advancing to the next row, when looking at the RIS from the front. Using this definition, Fig. \ref{fig:123} shows the log-magnitude and the phase of the reflect coefficients of the first three elements in the top left corner of the surface. The figure shows that the magnitude of each element takes roughly the same shape of a flat bell curve with its peak of around -7 dB being at $0^\circ$. This means that an angle-dependent attenuation is present, which increases up to -21.5 dBm when the relative angle towards the receiver is changed by $65^\circ$. As a result, an angle-dependent loss of up to -14.5 dB can be expected, depending on the relative angle of the receiver during the transmission.
The lower part of Fig. \ref{fig:123} shows the phase of the respective reflecting element. It can be clearly observed that the angle-dependent phase difference varies strongly between the elements, despite them being next to each other. More precisely, the phase deviation for element $m=1$ is $-0.01 \frac{\text{rad}}{^\circ}$, while the deviation for $m=2$ is $-0.04 \frac{\text{rad}}{^\circ}$, and for $m=3$ is $-0.09 \frac{\text{rad}}{^\circ}$. It seems that the closer the element is to the center column of the RIS, the stronger the phase deviation. To verify this theory, Fig. \ref{fig:141516} shows the elements in the top right row of the surface. Interestingly, the phase deviation of these elements also gets stronger, the closer the element is to the center of the RIS.  However, the signs of the deviations are inverted compared to the elements on the left.

\begin{figure}
\centering
\includegraphics[width=.975\linewidth]{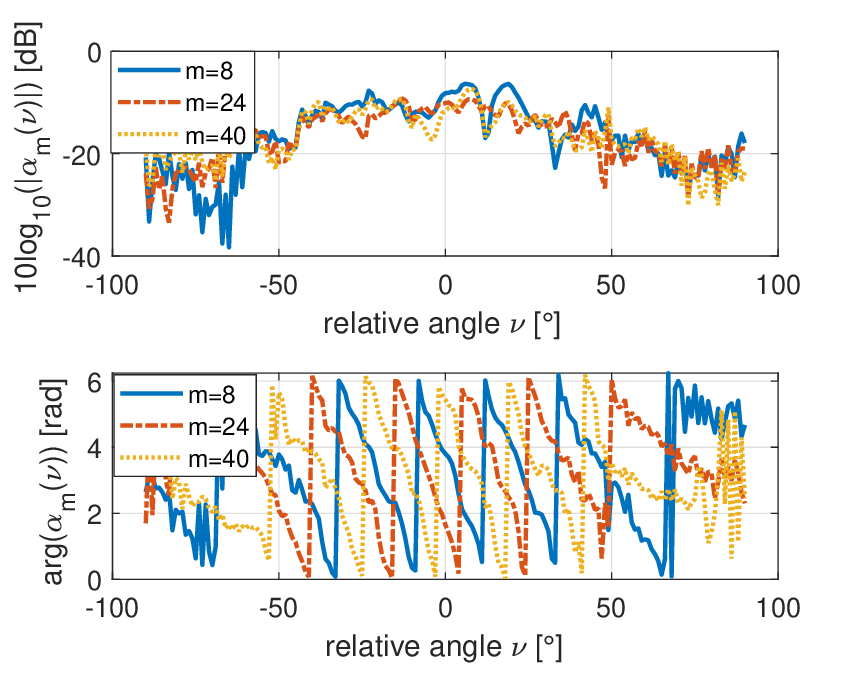}
\caption{Logarithmic magnitude (top) and phase (bottom) of the angle-dependent reflect coefficient $\alpha_m(\nu), \forall  m\in\{8,24,40\}$}
\label{fig:82440}
\end{figure}

\begin{figure}
\centering
\includegraphics[width=.975\linewidth]{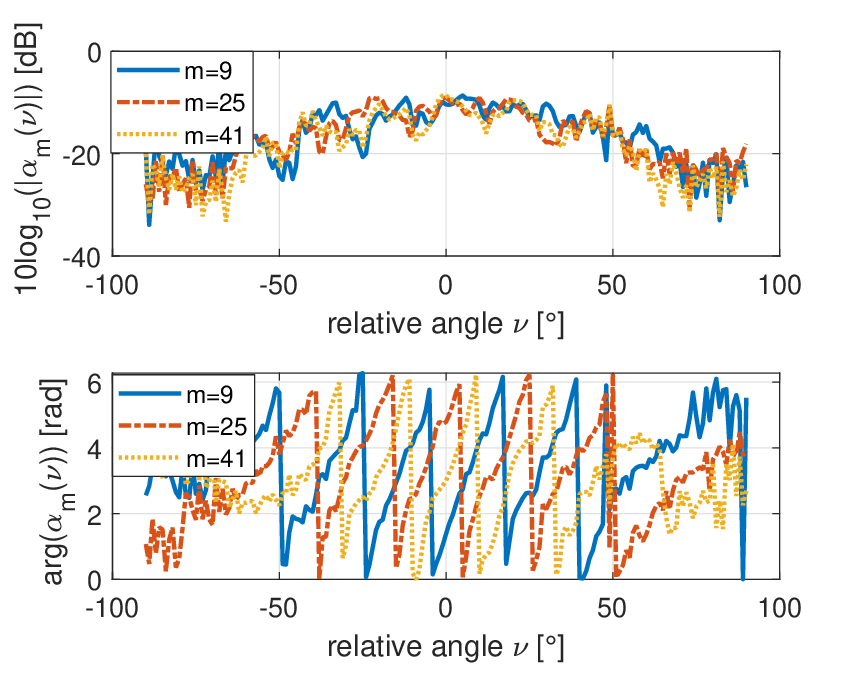}
\caption{Logarithmic magnitude (top) and phase (bottom) of the angle-dependent reflect coefficient $\alpha_m(\nu), \forall  m\in\{9,25,41\}$}
\label{fig:92541}\vspace{-0.4cm}
\end{figure}

As the surface is rotated vertically around its center, i.e., between column 8 and 9, we plot the reflection coefficients of the upper most elements within these two columns in Fig. \ref{fig:82440} and \ref{fig:92541}, respectively. These figures validate that the phase difference is strongest near the central columns of the surface with around $-0.285 \frac{rad}{^\circ}$ for the elements in the 8-th and $0.285 \frac{rad}{^\circ}$ for the 9-th column. It can be seen again that the phase deviation has a different sign between the left and right half of the surface. Notably, both figures show that the strength, i.e., slope, of the deviation does not change in the vertical direction. Further, the an offset in phase between the reflect coefficients can be observed. In conclusion, our findings emphasize the critical role of modeling the angle-dependent phase deviations, considering the consistent asymmetry between the left and right halves of the RIS.

\vspace{-0.15cm}

%
%


\section{Conclusion}
This work studies the deviation in attenuation and phase, that a RIS exhibits when reflecting the signal towards angles, which are not perpendicular to the RIS. To this end, a baseline channel model is utilized, which consists of the free-space path loss in combination with the attenuation that occurs during RIS switching. An optimization problem is formulated to obtain  RIS configurations that are optimal for beamsteering towards different angles. The determined configurations are measured in an anechoic chamber using a RIS prototype. By formulating a system of linear equations, we are able to extract the differences between simulated and measured values for each configuration by calculating the angle-dependent reflect coefficient for each RIS element. As it turns out, the precision of the simulations can be significantly improved to match the experimentally obtained measurements, if the angle-dependent coefficients are utilized during simulations. Upon analyzing the reflect coefficients, we highlight that the relative angle towards the receiver can induce an attenuation of up to -14.5 dB. Further, we are able to show that not only the magnitude, but also the phase is deviating, relative to angle towards the receiver. More precisely, the phase deviation of the outer rows of the surface increases towards the inner rows from $0.01\frac{rad}{^\circ}$ up to $0.285 \frac{rad}{^\circ}$, while the signs between the phase deviation on left and right side of the surface are inverted.
\section{Acknowledgement}
We would like to express our gratitude to M. Heinrichs
from the High Frequency Laboratory, TH
Cologne for providing his invaluable expertise during the experimental measurements.
\bibliographystyle{IEEEtran}
\bibliography{bibliography}
\balance
\end{document}